\documentclass{fizik}
\usepackage{hyperref}
\hypersetup{
colorlinks=true,
urlcolor=black,
citecolor=blue}
\usepackage[all]{xy,xypic}
\usepackage{amsfonts,amssymb,amsmath,amsgen,amsopn,amsbsy,theorem,graphicx,epsfig}
\usepackage{eufrak,amscd,bezier,latexsym,mathrsfs,enumerate,multirow}
\usepackage[utf8]{inputenc}
\usepackage[english]{babel}
\usepackage[dvipsnames]{xcolor}
\usepackage{academicons}
\definecolor{orcidlogocol}{HTML}{A6CE39}
\usepackage[pagewise]{lineno}
\usepackage{epstopdf}
\usepackage{xspace}
\usepackage{color}
\usepackage{units}
\usepackage{slashed}
\usepackage{gensymb}
\usepackage{booktabs}
\usepackage{xcolor}
\usepackage{setspace}
\usepackage{physics}
\usepackage{nicefrac}
\usepackage{bbold}
\usepackage{accents}
\usepackage{mathtools}
\usepackage{tikz}
\usepackage{amsmath}
\usepackage{amssymb}

\def\bt{\begin{equation}}
\def\bea{\begin{eqnarray}}
\def\ee{\end{equation}}
\def\eea{\end{eqnarray}}

\yil{}
\vol{}
\fpage{}
\lpage{}
\doi{}

\title{An analytical solution
of a quantum system with non-Markovian
behavior: The Bixon-Jortner system in the time domain
}

\author[Cevheroğlu and Özakın]{
	\textbf{Osman Cevheroğlu$^{1}$, Arkadaş Özakın$^{1}$\thanks{arkadas.ozakin@bogazici.edu.tr}}\\
    \\
	$^{1}$Department of Physics, Boğaziçi University
}

\newcommand{\bc}{\begin{center}}
\newcommand{\ec}{\end{center}}

\renewcommand{\phi}{\varphi}

\begin{document}

\maketitle

\begin{abstract} Non-Markovian behavior in quantum systems is often studied in bipartite systems consisting of a system of interest and an environment, where tracing over the environment results in non-Markovian dynamics for the subsystem of interest. In certain regimes, the dynamics has a Markovian limit described by the Lindblad master equation, and corrections to this behavior can be obtained by techniques such as the Nakajima-Zwanzig formalism. In this paper, we obtain an exact equation for the dynamics of a simple model system whose Hilbert space is a direct sum of a discrete state and an infinite ladder. This system, called the Bixon-Jortner model, originated in quantum chemistry but has been used in quantum optics as a model system with interesting behavior, including a Wigner-Weisskopf limit of exponential decay. We work directly in the time domain, starting from an integrodifferential equation for the amplitude of the discrete state. Using tools from mathematical physics, we transform this equation to a delay differential equation, which makes the memory effects completely transparent, and solve it using an intuitive ansatz. This gives the analytical form of the dynamics in the time domain, showing how decay and revivals arise from the delay structure. We also show that the model is equivalent to the one-excitation sector of a qubit coupled to a bosonic bath with a frequency-comb spectral density, so that the same solution determines the exact amplitude-damping dynamics of the qubit, and the revivals occur precisely where this dynamics fails to be CP-divisible. We verify our solution by direct numerical evolution of a truncated Hamiltonian. We believe the explicit form of the time-domain delay equation we obtain and the accessibility of the solution techniques we use make our results a useful case study of non-Markovianity in quantum systems.

\end{abstract}

\section{Introduction}
\label{section1} 
Decays and revivals are two archetypal dynamical 
behaviors observed in quantum optics and many other quantum systems. Decays most 
commonly occur in systems where a discrete set of states couple 
to a continuum; if the coupling varies slowly as a function of 
the continuum energy, then the initial discrete state ``leaks'' 
into the continuum in an irreversible manner with an 
exponential decay constant given by the Fermi 
golden rule. \cite{dirac1927quantum} In such cases, one often thinks of the 
continuum as an external environment that 
the discrete system of interest is coupled to.
While such irreversible time evolution does not appear unitary
when restricted to the subsystem of interest,
it is still Markovian, being described by a Lindblad master
equation with a time-independent generator, i.e., a quantum
dynamical semigroup.
(See \cite{breuer2002theory}
for a microscopic derivation of the master equation from
the Schr\"odinger equation under appropriate assumptions.)

Complete or partial \emph{revivals} occur when an initially decaying 
population of a state later starts increasing 
again, due to the leaked information finding its way back into the system of interest from the environment. 
This can happen when the environment itself is a discrete 
system or when the coupling 
between the system of interest and the environment is not slowly-varying.
When restricted to the subsystem of interest,
the dynamics in this case appears not only non-unitary,
but also non-Markovian: it can no longer be described by a
Lindblad equation with nonnegative decay rates, and fails to be CP-divisible (see Section~\ref{sec:tensor-embedding} for a precise statement in the setting of this paper).

While decay and revival dynamics are often observed in 
quantum optical systems, the fundamental light-matter interaction
describing such systems is complex, and analytical treatments are rare. For 
this reason, 
simple model systems that contain the essence of such behaviors are 
valuable tools for developing intuition and understanding,
which may then be applied to more realistic settings. 
The construction of useful model systems with all 
the ingredients of interest by stripping off the inessential 
mathematical complication is something of an art. Some of the pristine
examples of this art are the Weisskopf-Wigner model \cite{weisskopf1930berechnung},
the Jaynes-Cummings model \cite{jaynes1963comparison}, and the spin-boson model \cite{leggett1987dynamics}.

Originally developed in the quantum chemistry literature, the Bixon-Jortner (BJ) system \cite{bixon1968intramolecular}
is another simple model system that has exponential decay 
behavior as a limiting case, and decay-revival dynamics more 
generally. Its one fundamental difference from
the spin-boson and Jaynes-Cummings systems is that the full system 
is not a tensor product of a discrete system of interest with an 
environment, but is rather a \emph{direct sum} of two parts. Partly
for this reason, the system is simpler than the alternatives
while still displaying a range of interesting behaviors.
This simplicity makes the BJ system one of the 
often-quoted models in the quantum optics literature, 
studied in detail in textbooks such as 
\cite{barnett2002methods} and \cite{cohen1998atom}.

Simply put, the BJ system consists of a single discrete state denoted
by $|\varphi\rangle$, and an infinite ``ladder'' of states $|n\rangle$.
The unperturbed part $H_0$ of the total Hamiltonian 
is diagonal in this basis, with eigenvalues $E_{\varphi}$ and $E_n = n\Delta$
for $|\varphi\rangle$ and $|n\rangle$, respectively,
where $\Delta$ is the spacing of the ladder. The 
perturbation 
$V$ that provides the coupling
has the nonzero matrix elements $\langle \varphi|V|n\rangle = v$.
See Figure \ref{fig:energy_levels}
for a visual representation, and Section \ref{sec:the_model} for more detail.
\begin{figure}[h!]
    \centering
    \begin{tikzpicture}[
        scale=0.8, transform shape, 
        level/.style={draw=blue!60!black, very thick},
        desc_label/.style={text width=3.5cm, align=center},
        elabel/.style={anchor=west, font=\large}
    ]
        \def\spacing{0.45}

        \foreach \y in {0, 1, 2, 3, 4} {
            \pgfmathtruncatemacro{\n}{\y-2} 
            \draw[level] (1, \y*\spacing) -- (3, \y*\spacing); 
            \node[elabel] at (3.4, \y*\spacing) { 
                \ifnum\n=0 $0$\fi
                \ifnum\n=1 $\Delta$\fi
                \ifnum\n>1 $\n\Delta$\fi
                \ifnum\n=-1 \makebox[0pt][r]{$-$}$\Delta$\fi
                \ifnum\n<-1 \pgfmathtruncatemacro{\nabs}{abs(\n)}\makebox[0pt][r]{$-$}$\nabs\Delta$\fi
            };
        }
        \node[font=\Large] at (2, 5.0*\spacing) {$\vdots$}; 
        \node[font=\Large] at (2, -0.5*\spacing) {$\vdots$}; 
        \node[desc_label] at (2, -1.5) {Ladder states}; 

        \draw[level] (-3, 1.5*\spacing) -- (-1, 1.5*\spacing) node[midway, above, font=\large] {$E_{\varphi}$};
        \node[desc_label] at (-2, -1.5) {The ``discrete'' state};

    \end{tikzpicture}
    \caption{The Bixon-Jortner basis and unperturbed eigenvalues. The perturbation $V$ (not shown) couples $|\varphi\rangle$ to all the $|n\rangle$ with a constant coupling $v$.}
    \label{fig:energy_levels}
\end{figure}

There are two main approaches to the analysis of the Bixon-Jortner system:
the diagonalization of the full Hamiltonian, or a direct computation of 
its dynamics. In the former method, 
also called the ``dressed state'' approach in the
quantum optics literature, one starts by writing
the time-independent Schr\"odinger equation in the $|n\rangle$, $|\varphi\rangle$ basis, resulting in an infinite set of equations
for the eigenvectors and the eigenvalues.
These are then transformed by summing an infinite series using 
the Mittag-Leffler theorem, and one ends up with a transcendental equation
whose solutions give the eigenvalues and the eigenvectors. 
See \cite{barnett2002methods}
and \cite{cohen1998atom} for details. 

A direct approach to the computation of the dynamics is pursued 
in \cite{barnett2002methods},
where one obtains the Laplace transform of the coupled
ODEs representing the time-dependent Schr\"odinger equation with the initial state being the discrete
state, and then 
(after a bit of a tour-de-force) inverts the Laplace transform
to obtain the time-domain solution for the initial condition $|\psi(0)\rangle = |\varphi\rangle$.\footnote{Note that the result given in 
\cite{barnett2002methods} contains a small 
typo involving a sign error.}

Both of these approaches are valuable---one gives
analytical information on the spectrum of the Hamiltonian, and the other gives the full dynamics of the initial discrete state
in terms of special functions. However, from a physical point of view,
there is still something to be desired for a deeper understanding
of the behavior of this simple but interesting system.

\begin{figure}[htp]
    \centering
    \includegraphics[width=0.5\textwidth]{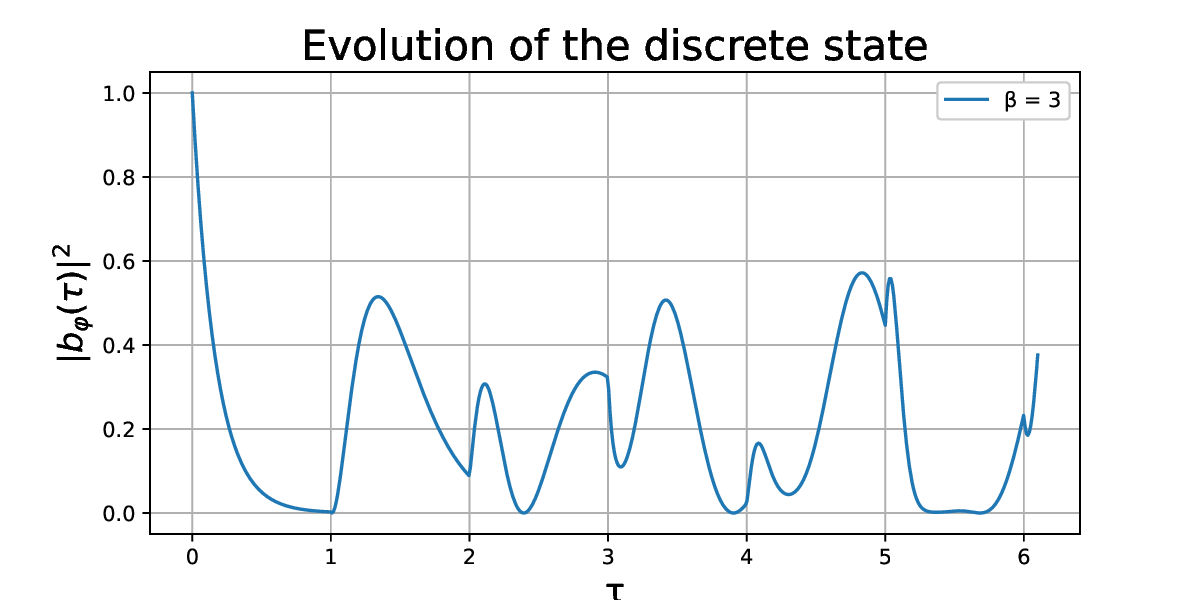}
    \caption{The time evolution of the probability
    of the discrete state in the BJ model as a 
    function of dimensionless time $\tau$, starting with
    the initial condition $b_{\varphi}(0)=1$, with $\beta=3$ and $\omega_\varphi=0$.
    Until $\tau=1$, the
    probability decays exponentially. Thereafter, it gets a ``kick'' at each integer time.}
    \label{fig:example-evolution}
\end{figure}

To get some feeling for the dynamics involved, let us look at 
a typical example of the time evolution of an initial discrete state.
In Figure \ref{fig:example-evolution}, we show the probability
$|b_{\varphi}|^2$ of 
$|\varphi\rangle$ as a function of time for a given choice of parameters.
The amplitude starts decaying 
exponentially, and then, at time $\tau=1$ (measured in appropriate units),
gets a kick, and the discrete-state population undergoes a partial revival. A smooth behavior
continues until $\tau=2$, at which time one gets another non-smooth kick.
At each integer time $\tau=n$, the system gets another kick. 
In other words, the behavior in Figure \ref{fig:example-evolution} starts as Markovian (an 
exponential decay), but then gets modified into a more complicated form, 
with additional complications at each integer time. The Laplace transform
approach of \cite{barnett2002methods} gives an analytical formula for the behavior, 
but the ``kicks'' appear somewhat out of the blue.
It would be helpful to have a more direct derivation 
for 
such significant qualitative aspects of the solution. 

In this paper, we attack the problem of the dynamics of the Bixon-Jortner 
system directly in the time domain. We start by writing the dynamics
of the discrete state in terms of an integro-differential equation, and 
then, using the Poisson sum formula, we turn the equation into a \emph{delay
differential equation}, where the time derivative of the coefficient 
$b_{\varphi}$ of the discrete state is given in terms of the values of 
$b_{\varphi}$ at previous instants, separated from the current instant by
integer multiples of a fundamental period.
This shows that the coefficient 
$b_{\varphi}(\tau)$ satisfies an equation that itself gets periodic kicks involving
previous 
values of the function $b_{\varphi}(\tau)$. We then
use a recurrence relation/generating function approach to solve this delay differential 
equation directly. The simplicity of the techniques
used make our solution an interesting and explicit 
case study of the appearance of non-Markovian dynamics in a model system.

The rest of the paper is organized as follows. In Section~\ref{sec:the_model},
we describe the model and show that it is equivalent to the one-excitation
sector of a qubit coupled to a bosonic bath. In Section~\ref{sec:the-equation}, we
derive the delay differential equation satisfied by the coefficient of the
discrete state, interpret its delay structure in terms of the frequency-comb
spectral density of the bath, and discuss the singular nature of the
constant-coupling limit (Section~\ref{sec:spectral-density}). In
Section~\ref{sec:the-solution}, we solve the delay equation using a recurrence
relation and a generating function, and verify the solution by direct
numerical evolution of a truncated Hamiltonian, which also gives the dynamics
of the ladder states (Section~\ref{sec:numerics}). In
Section~\ref{sec:discussion}, we discuss our results, possible physical
realizations of the model, and its limitations.

\section{The model}\label{sec:the_model}
\subsection{Description of the model}\label{sec:model-description}
The Hilbert space of the model is spanned by the state $|\varphi\rangle$ (the 
``discrete state'') and the infinite ladder of states $|n\rangle$, $n=0, \pm 1, 
\pm 2, \ldots$, which  
become a quasi-continuum of states in the limit of small spacing. The Hamiltonian is 
given by 
\begin{align}\label{eqn:bj-hamiltonian-simple}
H = H_0 + V   \,, 
\end{align}
where $H_0$ is diagonal in the $|n\rangle$, 
$|\varphi\rangle$ basis,
\begin{align}
H_0|n\rangle&=E_n|n\rangle  \\
H_0|\varphi\rangle&=E_{\varphi}|\varphi\rangle\,,
\end{align}
and $V$ couples the ladder to the discrete state, 
\begin{align}
\langle n|V|\varphi\rangle&=v = \overline{\langle \varphi|V|n\rangle}\\
\langle n|V|n\rangle&=0  = \langle \varphi|V|\varphi\rangle\,.
\end{align}
where $v$ is a constant. We will shortly take $E_n = n\Delta$, where $\Delta$ 
will be the uniform 
spacing of the ladder. We are interested in the dynamics 
of the system
when the initial state is the discrete state $|\Psi(0)\rangle = 
|\varphi\rangle$.

A direct approach to the dynamics can be pursued by
writing 
the time-dependent Schr\"odinger equation 
in the $|n\rangle$, $|\varphi\rangle$ basis, which gives an infinite 
set of coupled differential equations for the coefficients of 
these vectors. 
Defining,
\begin{equation}
  |\Psi(t)\rangle = c_{\varphi}(t)|\varphi\rangle+\sum_{n} c_n(t) |n \rangle,
\end{equation}
the initial conditions are $c_{\varphi}(0)=1$, $c_n(0)=0$. The Schrödinger equation,
\begin{equation}
i\hbar \frac{d}{dt}|\Psi\rangle = (H_0 + V)|\Psi\rangle\,,
\end{equation}
becomes, upon computing the inner products with
$| n\rangle$ and $| \varphi\rangle$, respectively,
\begin{align}    
 i\hbar \frac{dc_n}{dt} &= E_n c_n(t) + v c_\varphi(t) \\
 i\hbar\frac{dc_{\varphi}}{dt} &= E_{\varphi}c_{\varphi}(t) +\sum_n 
 \overline{v} c_n(t) \,.
\end{align}
Defining $\omega_n = E_n/\hbar$, $\omega_{\varphi} = E_{\varphi}/\hbar$ and using
$c_n(t) = b_n(t) e^{-i\omega_n t}$, $c_{\varphi}(t) = b_{\varphi}(t)e^{-
i\omega_{\varphi}t}$ to go to the interaction picture,
we get the Schrödinger equation in terms of the $b$ coefficients,
\begin{align}
\frac{db_n}{dt} &= -\frac{iv}{\hbar} e^{i(\omega_n-\omega_{\varphi})t}b_{\varphi}(t) \label{eqn:bn-sch}\\
\frac{db_{\varphi}}{dt} &= -\frac{i\overline{v}}{\hbar} \sum_n e^{i(\omega_{\varphi}-\omega_n)t}b_{n}(t) \,.\label{eqn:bphi-sch}
\end{align}

In the second chapter of  \cite{barnett2002methods}, the authors use
a Laplace transform on this set of equations
with the given initial condition to obtain
a set of algebraic equations, which is then solved 
for the Laplace transform of the coefficient of $b_{\varphi}$.
However, inverting this Laplace transform turns out to be
a bit involved, and is delegated to an Appendix in \cite{barnett2002methods}. 
The final result 
turns out to have ``kicks'' at each integer time (in 
appropriate units) where the behavior of the system
has non-smooth transitions.


\subsection{The tensor product embedding}\label{sec:tensor-embedding}
We next describe an embedding of the system described in Section \ref{sec:model-description}
into a bipartite system consisting of a qubit and a bath.
Consider a two-state system $S$ and a collection
of oscillators with frequencies $\omega_n$, which we call the bath, $B$.
The Hamiltonian of the combined system is taken to be
\begin{align}
    H = H_S + H_B + H_I \label{eqn:spin-boson-h}
\end{align}
where $H_S$ and $H_B$ are the system and bath Hamiltonians, respectively, given by\footnote{From this point on, we will work in units where $\hbar=1$, and assume the coupling $v$ is real---only the absolute value of $v$ enters our equations, so there is no loss of generality here.},
\begin{align}
    H_S &= \omega_0 \sigma_+\sigma_-\\
    H_B &= \sum_n\omega_na^{\dagger}_na_n\,,
\end{align}
and $H_I$ denotes the interaction Hamiltonian, which we take to be,
\begin{align}
    H_I = \sum_n (g_n\sigma_-a^{\dagger}_n + g_n^{*}\sigma_+a_n)\,.
\end{align}
This is a basic spin-boson interaction in the 
rotating wave approximation;
$\sigma_{\pm}=\frac{1}{2}(\sigma_1\pm i\sigma_2)$, and we leave the spectrum $\omega_n$ and
the coupling constants $g_n$ unspecified, for the moment. The Hilbert space of the 
full system consists of a tensor product, $\mathcal{H}=\mathcal{H}_S\otimes \mathcal{H}_B$,
where $\mathcal{H}_S$ is the two-state Hilbert space spanned by the ground state $|g\rangle_S$
and the excited state $|e\rangle_S$, and
$\mathcal{H}_B$ is spanned by the basis vectors 
given in terms of a sequence $m_n$ of integers
labeling the excitation number of each oscillator, $|\{m_n\}\rangle_{B}$.

The Hamiltonian \eqref{eqn:spin-boson-h} conserves the excitation number
$\sigma_+\sigma_- + \sum_n a^{\dagger}_na_n $. As a result, the subspace $\mathcal{H}_1$
of one-excitation states, i.e., 
\begin{align}
    \mathcal{H}_1 = \operatorname{span}\{|e\rangle_S\otimes|0\rangle_B\}\oplus
    \operatorname{span}\{|g\rangle_S\otimes |1_n\rangle_B\}
\end{align}
is invariant under the time evolution by the total Hamiltonian \eqref{eqn:spin-boson-h}.
Here, $|1_n\rangle_B$ denotes the state where all the oscillators except the $n$th 
one
are in the ground state, and the $n$th oscillator is in the first excited state.
This subspace maps to the Bixon-Jortner Hilbert space under 
$|\varphi\rangle\longleftrightarrow |e\rangle_S\otimes|0\rangle_B=|e, 0\rangle$
and $|n\rangle \longleftrightarrow |g\rangle_S\otimes|1_n\rangle_B=|g, 1_n\rangle$.
If we assume $\omega_n = n\Delta$, $n\in\mathbb{Z}$, $g_n=v$, 
$\omega_0 = E_{\varphi}$, the Hamiltonian \eqref{eqn:spin-boson-h}
of the spin-bath system becomes exactly the Bixon-Jortner Hamiltonian \eqref{eqn:bj-hamiltonian-simple}.\footnote{
The frequencies $\omega_n=n\Delta$, and likewise $\omega_0$, are to be understood as measured relative to a
common reference frequency, i.e., as detunings in a rotating frame, rather than as the literal frequencies of physical oscillators.}
Thus, the Bixon-Jortner system describes the 1-excitation subspace of
an appropriate spin-boson Hamiltonian.

With this identification in mind,
let us next consider an initial tensor product state $|\psi(0)\rangle$ 
where the qubit is in an arbitrary state and the bath is in the ground state,
\begin{align}
    |\psi(0)\rangle = (c_e|e\rangle_S+c_g|g\rangle_S)\otimes |0\rangle_B
\end{align}
In the interaction picture where states evolve under $H_I$, the state $|g, 0\rangle$
doesn't change over time since it is annihilated by both $a_n$ and $\sigma_-$,
and the evolution of the state $|e, 0\rangle$ is given 
by the evolution of the initial discrete state $|\varphi\rangle$ of the BJ system,
\begin{align}
    |\varphi\rangle \longrightarrow b_{\varphi}(t) |\varphi\rangle + \sum_n b_n(t)|n\rangle.
\end{align}
In terms of the tensor product embedding, this gives,
\begin{align}
    |\psi(t)\rangle = c_e\Big[b_{\varphi}(t)|e, 0\rangle+\sum_nb_n(t)|g, 1_n\rangle\Big] + c_g|g, 0\rangle\,.
\end{align}
Computing the density matrix of this state and tracing out the bath, we get the qubit 
density matrix,
\begin{align}\label{eqn:spin-boson-rho-evol}
    \rho_S(t)=\begin{pmatrix}
        |b_{\varphi}(t)|^2\rho_{ee}(0) & b_{\varphi}(t)\rho_{eg}(0)\\
        b^*_{\varphi}(t)\rho_{ge}(0) & \rho_{gg}(0) + [1-|b_{\varphi}(t)|^2]\rho_{ee}(0)
    \end{pmatrix}\,.
\end{align}
Thus, the time evolution of an initial discrete state in the BJ system gives
the time evolution of the spin-boson system with an initial tensor product state,
where the bath is in the ground state. We also note that 
the quantum channel $\mathcal{E}_t$ sending $\rho(0)$ to 
$\mathcal{E}_t(\rho(0)) = \rho(t)$ has the Kraus operators
\begin{align}
 K_0(t) &= |g\rangle\langle g| + b_{\varphi}(t)|e\rangle\langle e| \\  
 K_1(t) &= \sqrt{1-|b_{\varphi}(t)|^2}|g\rangle\langle e|\,.
\end{align}
The channel $\mathcal{E}$ given by \eqref{eqn:spin-boson-rho-evol} is known 
as an amplitude damping channel. 

The function $b_{\varphi}(t)$, to be computed exactly in Section~\ref{sec:the-solution},
therefore plays a double role: it is both the
Bixon-Jortner survival amplitude, and the single complex-valued function that
completely determines the reduced dynamics of a qubit coupled to a
frequency-comb-shaped reservoir. In particular, the non-monotonicity of
$|b_{\varphi}(t)|$ associated with the delayed revivals to be discussed in
Section~\ref{sec:the-solution} will have a direct operational meaning in the open-system
language. To see this consider the time-local master equation
\begin{align}\label{eqn:time-local-master}
    \dot{\rho}_S(t) = -\frac{i}{2}S(t)[\sigma_+\sigma_-, \rho_S(t)] + \gamma(t)\big(\sigma_-\rho_S(t)\sigma_+-
    \frac{1}{2}\{\sigma_+\sigma_-, \rho_S(t)\}\big)
\end{align}
Using \eqref{eqn:spin-boson-rho-evol}, we get
\begin{align}\label{eqn:gamma-s-for-time-local}
    \gamma(t) = -2\operatorname{Re}\frac{
    \dot{b}_{\varphi}(t)}{b_{\varphi}(t)}\,,\quad{}
    S(t) = -2\operatorname{Im}\frac{\dot{b}_{\varphi}(t)}{b_{\varphi}(t)}\,.
\end{align}
Now, a process is CP-divisible 
\cite{rivas2010entanglement, breuer2016colloquium}
if and only if $\gamma(t)\ge 0$, which, according to
\eqref{eqn:gamma-s-for-time-local}, is equivalent to
\begin{align}\label{eqn:cp-divisibility-from-b}
    \frac{d}{dt}|b_{\varphi}(t)| \le 0\,.
\end{align}
Thus, the evolution of the system \eqref{eqn:time-local-master} is CP-divisible
if and only if the absolute value of the discrete state amplitude 
is decreasing (on intervals where $b_{\varphi}\neq 0$). 
In other words, when the absolute value of $b_{\varphi}$
is increasing, we can conclude that the process is non-CP-divisible,
and hence, non-Markovian in the sense of CP-divisibility. See \cite{lonigro2023excitation}
for a discussion of CP-divisibility of 
excitation-damping channels, and \cite{chruscinski2023markovianity}
for CP-divisibility in the context of multilevel spin--boson models.
(Note that
there exist stronger notions of non-Markovianity involving
multitime/process tensor approaches \cite{burgarth2021elusive}.) 

\section{The equation for $b_{\varphi}$}\label{sec:the-equation}
In this section, we obtain a delay differential equation satisfied by 
$b_{\varphi}$.

\subsection{An integrodifferential equation}
Equation \eqref{eqn:bn-sch} can be integrated immediately to get
\begin{equation}
 b_n(t) = -{iv}\int^{t}_0e^{i(\omega_n-\omega_{\varphi})t'} b_\varphi(t')dt'\,,
\end{equation}
where we used the initial condition $b_n(0)=0$. Plugging this in \eqref{eqn:bphi-sch}, we get
\begin{equation}
    \frac{db_{\varphi}}{dt} = -v^2 \sum^\infty_{n=-\infty}\int^{t}_0e^{i(\omega_n-\omega_{\varphi})(t'-t)} b_{\varphi}(t')dt'\,.
\end{equation}
This equation represents the dynamics of the coefficient 
of the initial state $|\varphi\rangle$---we got rid of the 
coefficients of the ladder states at the expense of turning the
coupled set of differential equations 
to an 
\emph{integrodifferential} equation for $b_{\varphi}$.
The derivative of $b_{\varphi}$ at a given instant
is now given in terms of its values at previous instants; the effects of the ladder are encoded in the integral and the sum.

Two commonly-used approximations for an equation of this form
are: (1) the first order perturbation theory approximation where one 
replaces $b_{\varphi}(t)$ on the right hand side with its initial 
value 
$b_{\varphi}(t')=1$, and (2) the Markovian approximation where one 
uses the final value $b_{\varphi}(t') = b_{\varphi}(t)$, turning the
equation to a differential equation.  The
latter approximation is analogous to the approach used in the 
Weisskopf-Wigner solution \cite{weisskopf1930berechnung} to the decay of a discrete state
into a continuum,
and likewise results in an exponential decay with a rate
given by the Fermi golden rule. The former approximation gives the
initial, linear part of this exponential decay. For the regimes 
of validity of these approximations, see, e.g., \cite{cohen1986quantum}.

We next specialize to the case of the Bixon-Jortner model by choosing $E_n = \hbar \omega_n = n\Delta$. This gives,
\begin{equation}
    \frac{db_{\varphi}}{dt} =-v^2 \sum^\infty_{n=-\infty}
    \int^{t}_0 e^{-i\omega_\varphi(t'-t)} e^{i n \Delta (t'-t)} 
    b_{\varphi}(t')dt'\label{eq:b-general-eq}
\end{equation}
Our aim here
is to solve \eqref{eq:b-general-eq} analytically and directly, 
without any approximations. This  
will give the full dynamics of the state $|\varphi\rangle$, and
the direct approach will make transparent the non-Markovian behavior 
resulting from the interaction with the ladder.

\subsection{Transforming to a delay differential equation}
We start by using the Poisson summation formula, \cite{stein2011fourier}
\begin{equation}\label{eq:poisson}
    \sum^\infty_{n=-\infty}e^{2\pi inx}=\sum^\infty_{m=-\infty}\delta(x-m) \,.
\end{equation}
Using this in \eqref{eq:b-general-eq} with $x=\Delta(t-t')/2\pi$, we get,
\begin{align}
    \frac{db_\varphi}{dt} 
    &=-\frac{2\pi v^2}{\Delta}  \sum^\infty_{m=-\infty}\int^{t}_0e^{-i\omega_\varphi(t'-t)}\delta\left(t'-[t-m t_0]\right )b_\varphi(t')dt' \label{eq:delta-ode}\,,
\end{align}
where $t_0=2\pi /\Delta$ is the time scale determined by the spacing 
of the ladder.
For $m=0$, the integral in the sum becomes,
\begin{align}
    \int^{t}_0  e^{-i\omega_\varphi(t'-t)}\delta\left(t'-t\right )b_\varphi(t')dt'=\frac{1}{2}b_\varphi(t)\,.
\end{align}
For $m\ne 0$, we have,
\begin{align}
    \int^{t}_0  e^{-i\omega_\varphi(t'-t)}\delta\left(t'-[t-mt_0]\right )b_\varphi(t')dt'=\Theta(t-mt_0)\Theta(mt_0)e^{im\omega_\varphi t_0}b_\varphi(t-mt_0)
\end{align}
where $\Theta(x)$ denotes the Heaviside step function defined as,
\begin{align}
\Theta(x) = \begin{cases}
    1 & \text{if } x>0,\\
    1/2 & \text{if } x=0,\\
    0 & \text{if } x<0\,.
    \end{cases}
\end{align}
Defining 
\begin{align}\label{eq:tau-t-definition}
 \tau=t/t_0 = \frac{t \Delta}{2\pi}
\end{align}
\eqref{eq:delta-ode} thus becomes,\footnote{
The $m=0$ delta function sits at the upper endpoint of the integration
range and contributes only half its weight, whereas for $t>mt_0$ the
$m\ge1$ terms lie strictly inside $(0,t)$ and contribute fully; this is
the origin of the relative factor of 2 in \eqref{eq:non-markov-ode}.}
\begin{align}\label{eq:non-markov-ode}
    \frac{db_\varphi}{d\tau}(\tau)&=-\beta\left[b_\varphi(\tau)+2\sum^\infty_{m=1}\Theta(\tau-m)e^{im\omega_\varphi t_0}b_\varphi(\tau-m)\right]\,,
\end{align}
where, 
\begin{align}
    \beta=\frac{2\pi^2 v^2}{\Delta^2}\,.\label{eq:beta-defn} 
\end{align}

\paragraph{Structure of the delay term.}
The first term on
the right hand side of \eqref{eq:non-markov-ode} 
is the instantaneous, local-in-time contribution, and would give a simple exponential
decay on its own, if one were to ignore the contribution from the sum.
That is, the differential equation becomes
\begin{align}
    \frac{db_\varphi}{d\tau}(\tau)=-\beta b_{\varphi}(\tau),
\end{align}
which has a simple solution
\begin{align}\label{b-for-first-inertval}
    b_\varphi(\tau)=e^{-\beta\tau}\quad 0<\tau<1,
\end{align}
for initial condition $b_\varphi(0)=1$.
In this period,
the decay constant for $|b_\varphi(t)|^2 = |b_\varphi(0)|^2e^{-\Gamma t}$ would be 
\begin{equation}\label{eq:gamma-fermi}
  \Gamma =  \frac{2 \beta}{t_0} =\frac{2\pi v^2}{\Delta} \,,
\end{equation}
which is exactly what one would get from the Fermi golden rule, since $\Delta$
is the inverse of the density of states. Note that in this interval, the 
CP-divisibility criterion \eqref{eqn:cp-divisibility-from-b} would
conclude that the dynamics is Markovian. 
This is closely related to the notion of ``hidden
non-Markovianity'' discussed in \cite{burgarth2021elusive,burgarth2021hidden},
where dynamics can be exactly Markovian for a
finite interval and reveal memory only later.

The sum in \eqref{eq:non-markov-ode} encodes the memory effects responsible for the loss of
CP-divisibility discussed in Section~\ref{sec:tensor-embedding}; it has contributions from the
past history of $b$. Up to $\tau=1$,
the equation has the exponential decay
behavior mentioned. At $\tau=1$, the $1$-delayed form of $b$ starts
contributing, and for $\tau\in (1, 2)$, the equation becomes,
\begin{equation}
     \frac{db_\varphi}{d\tau}(\tau)=-\beta\left[b_\varphi(\tau)+2e^{i\omega_\varphi t_0}b_\varphi(\tau-1)\right]\,.\label{eq:non-markov-2-term}
\end{equation}
The derivative at time $\tau$ is now given in terms of the values of the function at both
$\tau$ and $\tau-1$, reflecting the entry of the first delayed contribution.
As a result, the behavior starts to differ
from a simple exponential decay in this interval.  
To solve this differential equation in the interval $1<\tau<2$, we note that the
argument $\tau-1$ of the delayed term then lies in $(0,1)$, where
\eqref{b-for-first-inertval} applies. Thus, we can write
\begin{align}\label{shifted-b-for-first-kick}
    b_\varphi(\tau-1)=e^{-\beta(\tau-1)} \quad 1<\tau<2\,.
\end{align}
Inserting equation \eqref{shifted-b-for-first-kick} into \eqref{eq:non-markov-2-term} we get
\begin{align}
    \frac{db_\varphi}{d\tau}(\tau)=-\beta\left[b_\varphi(\tau)+2e^{i\omega_\varphi t_0}e^{-\beta(\tau-1)}\right]
\end{align}
Multiplying with $e^{\beta \tau}$ and rearranging gives
\begin{align}
    \frac{d}{d\tau}(e^{\beta\tau}b_\varphi(\tau))=-2\beta e^\beta e^{i\omega_\varphi t_0}\,,
\end{align}
which has a general solution,
\begin{align}
    b_\varphi=Ce^{-\beta \tau} -2\beta \tau e^{i\omega_\varphi t_0}e^{-\beta(\tau-1)}\,.
\end{align}
Since $b_\varphi(\tau)$ is continuous, from equation \eqref{b-for-first-inertval} we can get the initial condition for this interval as
\begin{align}
    b_\varphi(1)=e^{-\beta}.
\end{align}
Thus, $C=1+2\beta e^\beta e^{i\omega_\varphi t_0}$ and the solution can be written as
\begin{align}
    b_\varphi(\tau)=e^{-\beta\tau}[1-2\beta e^\beta  e^{i\omega_\varphi t_0}(\tau -1)].
\end{align}

Similarly, at $\tau=2$, the $2$-delayed
form of $b_{\varphi}$ starts contributing, and for $\tau\in (2, 3)$ the equation becomes,
\begin{equation}
     \frac{db_\varphi}{d\tau}(\tau)=-\beta\left[b_\varphi(\tau)+2(e^{i\omega_\varphi t_0}b_\varphi(\tau-1)+e^{2i\omega_\varphi t_0}b_\varphi(\tau-2))\right]\,.\label{eq:non-markov-three-term}
\end{equation}
In this way, at each integer value of $\tau$,
the behavior of $b_{\varphi}$ starts to see the effect of a new delayed version. 
Equations such as \eqref{eq:non-markov-ode}, \eqref{eq:non-markov-2-term}, \eqref{eq:non-markov-three-term}, where the derivative at a given time 
is given in terms of the values at certain preceding times are called delay differential equations
\cite{bellman1963differential}. 
While there are some techniques for dealing with such equations, there is 
no general method. 

\subsection{Spectral-density interpretation}\label{sec:spectral-density}

Equation~\eqref{eq:b-general-eq} can be recast in the standard language of open quantum
systems~\cite{breuer2002theory}, where the reduced dynamics of a system coupled to a bath is
written in terms of a bath correlation function $C(u)$,
\begin{align}\label{eq:corr-function}
\frac{db_\varphi}{dt}(t) = -\int_0^t C(t-t')\, b_\varphi(t')\, dt', \qquad
C(u) = \sum_n |g_n|^2 e^{-i(\omega_n - \omega_\varphi) u}
= \int \frac{d\omega}{2\pi}\, J(\omega)\, e^{-i(\omega-\omega_\varphi) u}\,,
\end{align}
where $J(\omega) = 2\pi \sum_n |g_n|^2 \delta(\omega - \omega_n)$ is the bath spectral density.
For the Bixon-Jortner choice $g_n = v$, $\omega_n = n\Delta$, the spectral density is not a smooth
function of $\omega$ but an ideal, infinitely sharp frequency comb,
\begin{align}\label{eq:spectral-density}
J_{\mathrm{BJ}}(\omega) = 2\pi v^2 \sum_{n \in \mathbb{Z}} \delta(\omega - n\Delta)\,.
\end{align}
The Fourier transform of $J_{\mathrm{BJ}}$ can be obtained by the same Poisson summation
formula~\eqref{eq:poisson} already used above, now applied directly to the correlation
function rather than to the equation of motion:
\begin{align}\label{eq:corr-poisson}
C(u) = v^2 \sum_n e^{-i(n\Delta - \omega_\varphi)u}
= \frac{2\pi v^2}{\Delta} \sum_{m \in \mathbb{Z}} e^{i m \omega_\varphi t_0}\, \delta(u - m t_0)\,.
\end{align}
Thus $C(u)$ is itself a Dirac comb in time, with teeth located precisely at the recurrence times
$u = m t_0$: the $m=0$ tooth reproduces the instantaneous (Markovian) term in
Eq.~\eqref{eq:non-markov-ode}, while the $m \geq 1$ teeth --- with the phases $e^{i m \omega_\varphi t_0}$ already appearing in our delayed terms --- are responsible for the periodic
kicks. This gives a direct physical interpretation of the delay structure: it is the temporal
image of an idealized comb-shaped reservoir spectrum.

The same picture clarifies the origin of the golden-rule rate $\Gamma$. Coarse-graining
$J_{\mathrm{BJ}}(\omega)$ over many teeth, i.e., treating the bath as effectively smooth on timescales short compared to $t_0$, replaces the comb by its local average,
\begin{align}
J_{\mathrm{BJ}}(\omega_\varphi) \;\longrightarrow\; \frac{2\pi v^2}{\Delta} = \Gamma\,,
\end{align}
recovering Eq.~\eqref{eq:gamma-fermi}. In this sense the unresolved comb looks like a flat
Markovian reservoir at short times; only once the individual teeth are resolved, at $t \gtrsim
t_0$, does the exact comb structure, and with it the revival dynamics, become visible. The
emergence and breakdown of such Markovian descriptions from an underlying oscillator-bath spectral density is discussed in detail in~\cite{santra2017fermi}.


A caveat is in order regarding the idealized coupling $g_n = v$ of the Bixon-Jortner system.
An interaction term of the form
$H_I = \sigma_- a^{\dagger}(g) + \sigma_+ a(g)$, $a^{\dagger}(g) = \sum_n g_n a_n^{\dagger}$,
normally requires a square-summable form factor, $\sum_n |g_n|^2 < \infty$. The constant
coupling $g_n = v$ over the two-sided infinite ladder violates this,
$\sum_{n\in\mathbb{Z}}|g_n|^2 = \infty$, so the Bixon-Jortner Hamiltonian, in this strict
sense, describes a singular idealization rather than an ordinary, bounded interaction. This is
reflected in the short-time behavior of our solution: a square-summable
coupling gives the usual quadratic short-time survival law, $P_e(t) = 1-(\delta H)^2t^2+o(t^2)$,
whereas~\eqref{b-for-first-inertval} gives $|b_\varphi(\tau)|=e^{-\beta\tau}$, i.e.\ a
nonzero one-sided derivative at $\tau=0$. This  signature of the singular coupling
disappears once the coupling is regularized, for instance by a symmetric finite cutoff
$|n|\le N$ followed by $N\to\infty$. For finite 
$N$ (and for $\omega_{\varphi}=0$), the correlation function becomes a finite-width
pulse train,
\begin{align}
C_N(t) = v^2\sum_{n=-N}^{N} e^{-in\Delta t} = v^2\,\frac{\sin\!\big[(N+\tfrac12)\Delta t\big]}{\sin(\Delta t/2)}\,,
\end{align}
rather than a genuine delta comb, restoring the ordinary quadratic short-time law. The delay
equation~\eqref{eq:non-markov-ode} is recovered once the $N\to\infty$ limit is taken.
Alternatively, one may replace the constant coupling by an envelope $g_n\to 0$ as
$|n|\to\infty$. In \cite{isgorur2026coupling}, we pursue exactly this route
using a Lorentzian coupling profile, obtaining a regularized model
that interpolates between the decay-revival dynamics of the BJ system and Rabi oscillations. Standard
textbook treatments of the Bixon-Jortner system, including~\cite{cohen1998atom}, work with
the constant coupling directly and do not address this idealization. A fully rigorous
construction of such singular couplings, via generalized Friedrichs-Lee or spin-boson
Hamiltonians with non-normalizable form factors, is given in
\cite{facchi2021spectral,lonigro2022generalized}.

We will next use a simple, intuitive ansatz to solve \eqref{eq:non-markov-ode}.

\section{The solution}\label{sec:the-solution}
\subsection{Obtaining a recurrence relation}

To solve \eqref{eq:non-markov-ode}, we additively decompose the unknown 
function and obtain a recurrence relation between the terms.
Let us first define $f(\tau) = b_{\varphi}(\tau)e^{\beta \tau}$, which turns 
\eqref{eq:non-markov-ode} into,
\begin{equation}
\frac{df}{d\tau}(\tau)=
-2\beta\sum^\infty_{m=1}e^{\eta m}f(\tau-m)\Theta(\tau-m)\label{eq:f-version-ode}\,,
\end{equation}
where
\begin{align}
    \eta=\beta +i\omega_\varphi t_0\,.
\end{align}
Note that $\omega_{\varphi}$ enters \eqref{eq:f-version-ode}
only through $e^{2\pi i\omega_\varphi/\Delta}$, and so the
dynamics depends on $\omega_{\varphi}$ only modulo $\Delta$,
which is expected since shifting $E_{\varphi}$ by $\Delta$
simply relabels the ladder states.

We would like to solve \eqref{eq:f-version-ode} with the initial condition 
$b_{\varphi}(0) = f(0)=1$. For $0\le \tau <1$, the right hand side (RHS) is zero, and the solution is just the constant
function $f(\tau)=1$. 
As mentioned above, at each integer value of $\tau$, the equation 
gets modified and we get an additional delay term, modifying the behavior
of the solution. With this motivation, we will assume that
the solution for $\tau\ge 0$ can be written as,
\begin{equation}\label{eq:f-ansatz}
    f(\tau) = \sum^{\infty}_{n=0}f_n(\tau-n)\Theta(\tau-n)\,,
\end{equation}
which says that at each integer $\tau$,
a new additive piece $f_n$ of the solution ``kicks in''.
We will assume each $f_n(\tau)$ of this ansatz
to be a smooth function defined for all $\tau\ge 0$.

To solve for $f_n$, we pick an integer $k > 0$ and focus on the
time interval $\tau \in (k\,,k+1)$. 
In this interval, each step function in \eqref{eq:f-ansatz} and \eqref{eq:f-version-ode} can be replaced with its constant value\footnote{We will assume $f_n(\tau)$'s are continuous. Then, due to continuity of the state function it can be shown that $f_n(0)=0$ for $n\neq0$. It follows that choosing the value of the Heaviside function $\Theta(0)$ does not change the calculations.},
\begin{align}
\Theta(\tau-m) = \begin{cases}
    1 & \text{if } 0\le m\le k\\
    0 & \text{if } m>k\,.
    \end{cases}
\end{align}
We thus truncate the infinite sum in the differential equation \eqref{eq:f-version-ode} to get,
\begin{equation}
    \frac{df}{d\tau}(\tau) =
-2\beta\sum^k_{m=1}e^{\eta m}f(\tau-m)\,,\label{eq:k-f-version-ode}
\end{equation}
and the ansatz $\eqref{eq:f-ansatz}$ becomes,
\begin{equation}\label{eq:f-ansatz-finite}
f(\tau)=\sum^k_{n=0}f_n(\tau-n) \,.
\end{equation}
Since \eqref{eq:k-f-version-ode} involves $f(\tau-m)$ where the
argument of $f$ is shifted to the left by $m$, and thus is
in the interval $(k-m, k-m+1)$, we  note
\begin{equation}
    f(\tau-m) = \sum^{k-m}_{n=0}f_n(\tau-m-n)\,.\label{eq:f-ansatz-finite-shifted}
\end{equation}
Using \eqref{eq:f-ansatz-finite} and \eqref{eq:f-ansatz-finite-shifted}
in \eqref{eq:k-f-version-ode}, we get,
\begin{align}\label{eq:k-time-f}
\sum^k_{n=0}\frac{df_n}{d\tau}(\tau-n) =-2\beta\sum_{m=1}^k 
e^{\eta m}\sum_{n=0}^{k-m}f_n(\tau-m-n)\,.
\end{align}
This is an equation satisfied by the $f_n$, obtained
from the interval $\tau\in (k\,, k+1)$. Writing the 
corresponding equation for the 
previous interval, $\tau \in (k-1\,,k)$, we get,
\begin{align}\label{eq:k-minus-one-time-f}
\sum^{k-1}_{n=0}\frac{df_n}{d\tau}(\tau-n)=-2\beta\sum^{k-1}_{m=1}e^{\eta m}\sum^{k-1-m}_{n=0}f_n(\tau-m-n)
\end{align} 
Subtracting \eqref{eq:k-minus-one-time-f} from \eqref{eq:k-time-f},
we get,
\begin{align}
    \frac{df_k}{d\tau}(\tau-k)
     &=-2\beta\sum^{k-1}_{m=1}e^{\eta m}f_{k-m}(\tau-k)-2\beta e^{\eta k}f_0(\tau-k)\notag\,.
\end{align}
Absorbing the last term in the sum and replacing $\tau-k$ everywhere with $\tau$,
we get a simple equation satisfied by the $f_n$,
\begin{equation}\label{eq:f-req-1}
    \frac{df_k}{d\tau}(\tau)
    =-2\beta\sum^{k}_{m=1}e^{\eta m}f_{k-m}(\tau)\,.
\end{equation}
We further simplify by 
using another step of recurrence: we write the same equation for $k-1$,
\begin{align}
\frac{df_{k-1}}{d\tau}(\tau)&=-2\beta\sum^{k-1}_{m=1}e^{\eta m}f_{k-1-m}(\tau)\,,
\end{align}
which gives,
\begin{equation}\label{eq:f-req-2}
e^{\eta}\frac{df_{k-1}}{d\tau}(\tau)=
-2\beta \sum_{m=2}^{k}e^{\eta m}f_{k-m}(\tau)\,.
\end{equation}
Subtracting \eqref{eq:f-req-2} from \eqref{eq:f-req-1}, we get
the recurrence relation,
\begin{equation}
    \frac{df_k}{d\tau}(\tau) - e^{\eta} \frac{df_{k-1}}{d\tau}(\tau)
    =-2\beta e^\eta f_{k-1}(\tau)\,.
\end{equation}
Finally, defining,
\begin{align}\label{eq:h-f-T-tau-defn}
    \tilde{f}_k(\tau) &= f_k(\tau)e^{-\eta k}, \qquad{} 
    h_k(T) = \tilde{f}_k(\tau(T))
\end{align}
where $T=2\beta \tau$, we obtain the simplified 
recurrence relation,
\begin{align}\label{eq:h-recursion}
h_k'(T)-h_{k-1}'(T)= - h_{k-1}(T)\,.
\end{align}

\subsection{Solving the recurrence relation}

We will solve the recurrence relation \eqref{eq:h-recursion} using a generating
function approach. Let $g(T,s)$ be the generating function of $h_n$,
\begin{equation}
g(T,s)=\sum_{n=0}^{\infty} s^n h_n(T) \,.\label{eq:gen-fn-defn}
\end{equation}
Computing the partial derivative with respect to $T$ and
using the recurrence relation \eqref{eq:h-recursion} together with the fact
that $h_0$ is a constant, 
we get,
\begin{align}
   \frac{\partial g}{\partial T}(T,s) &=\sum^\infty_{n=1}s^nh_n' \\
    &=\sum^\infty_{n=1}s^n\left[ -h_{n-1}+h_{n-1}' \right] \\
    &=-s\sum^\infty_{n=0}s^{n}h_{n}+s\sum^\infty_{n=0}s^{n}h_{n}'  \\
    &=-sg(T,s)+s\frac{dg}{dT}(T,s)\,,
\end{align}
which gives,
\begin{align}
    \frac{\partial g}{\partial T}&=\frac{-s}{1-s}g \,.
\end{align}
This can be integrated to give
\begin{equation} \label{eq:gen-fn-pre-soln}
    g(T,s)=A(s)\exp\left(\frac{sT}{s-1}\right) \,,
\end{equation}
where $A(s)$ is an unknown function.
To determine  $A(s)$, consider $g(T=0,s)$: 
\eqref{eq:gen-fn-pre-soln} gives $g(0,s)=A(s)$, and
\eqref{eq:gen-fn-defn} gives $g(T=0,s)=\sum_{n=0}^{\infty} s^nh_n(0)$. The initial conditions give $h_0(0)=1$,
and the continuity of the state as a function of time
requires $h_n(0)=0$ for $n\ge 1$. This gives $A(s)=1$.
We get the final form of the generating function,
\[
g(T,s)=\exp\left(\frac{-sT}{1-s}\right)\,.
\]
\textbf{Relating to Laguerre polynomials.} This is very similar to the generating function $\ell(T,s)$ of the Laguerre
polynomials $L_n(T)$:
\begin{align}
    \ell(T, s) &= \sum_{n=0}^{\infty}s^n L_n(T) \\
    &= \frac{1}{1-s}\exp\left(\frac{-sT}{1-s}\right) = 
    \frac{1}{1-s}g(T,s)\,.
\end{align}
We thus get,
\begin{align}
g(s, T) &= 
\sum_{n=0}^{\infty} s^n h_n(T)\\
&= \exp\left(\frac{-sT}{1-s}\right)\\
&= (1-s)\ell(T, s)\\
&= (1-s)\sum_{n=0}^{\infty}s^n L_n(T)\\
&= 1 + \sum_{n=1}^{\infty}s^n (L_n(T) - L_{n-1}(T))
\end{align}
Equating the first and the last lines gives the solution
for all $h_n$:
\begin{align}
h_0(T) &= 1 = L_0(T)\\
h_n(T) &= L_n(T) - L_{n-1}(T) \qquad{} (n\ge 1)\,.
\end{align}
Using \eqref{eq:h-f-T-tau-defn}, we get the solution for $f_n$,
\[
f_n(\tau)=e^{\eta n}\left[L_n(2\beta \tau)-L_{n-1}(2\beta \tau)\right]\,.
\]
\subsection{The full solution}
Combining the results above, the full solution is given as,
\begin{align}
    b_\varphi(\tau) &= e^{-\beta \tau}f(\tau)\\
    &= e^{-\beta \tau}\sum_{n=0}^{\infty} \Theta(\tau-n)f_n(\tau-n)\\
    &= e^{-\beta \tau}\Bigl(1 + \sum_{n=1}^{\infty}e^{\eta n}\Theta(\tau-n) \bigl[L_n(2\beta (\tau-n)) \notag \\
    &- L_{n-1}(2\beta (\tau-n))\bigr]\Bigr)\,.\label{eq:our-form-solution}
\end{align}
Alternatively, it is
possible to use the recurrence relation of the Laguerre polynomials
\begin{equation}
 x L_n'(x) = n(L_n(x) - L_{n-1}(x))\,,
\end{equation}
to get, 
\begin{equation}
    b_\varphi(\tau) = e^{-\beta \tau}\left(1 + 
    2\beta \sum_{n=1}^{\infty}\Theta(\tau-n) e^{\eta n}
    \frac{\tau -n}{n}L_n'(2\beta(\tau-n))
    \right)\label{eq:barnett-radmore-form-solution}\,,
\end{equation}
    which is to be compared with the solution (2.5.8)
    given on page 28 of \cite{barnett2002methods}. For $\omega_\varphi=0$, the
    two results agree except for a sign difference in front of the sum term;
    Professor Barnett kindly confirmed via email that the sign in \cite{barnett2002methods}
    is a typo, and the direct numerical evolution of Section~\ref{sec:numerics}
    (Figure~\ref{fig:numerical-convergence}) confirms the sign
    in~\eqref{eq:our-form-solution}, 

Finally, using \eqref{eq:tau-t-definition},
one can get the solution in terms of real time by substituting
$\beta \tau = \Gamma t/2$, where, 
\begin{equation}
     \Gamma = \frac{2\pi v^2}{\Delta}
\end{equation}
is once again the decay constant one would get from the Fermi golden rule. 

\begin{figure}[htp]
    \centering
    \includegraphics[width=0.7\textwidth]{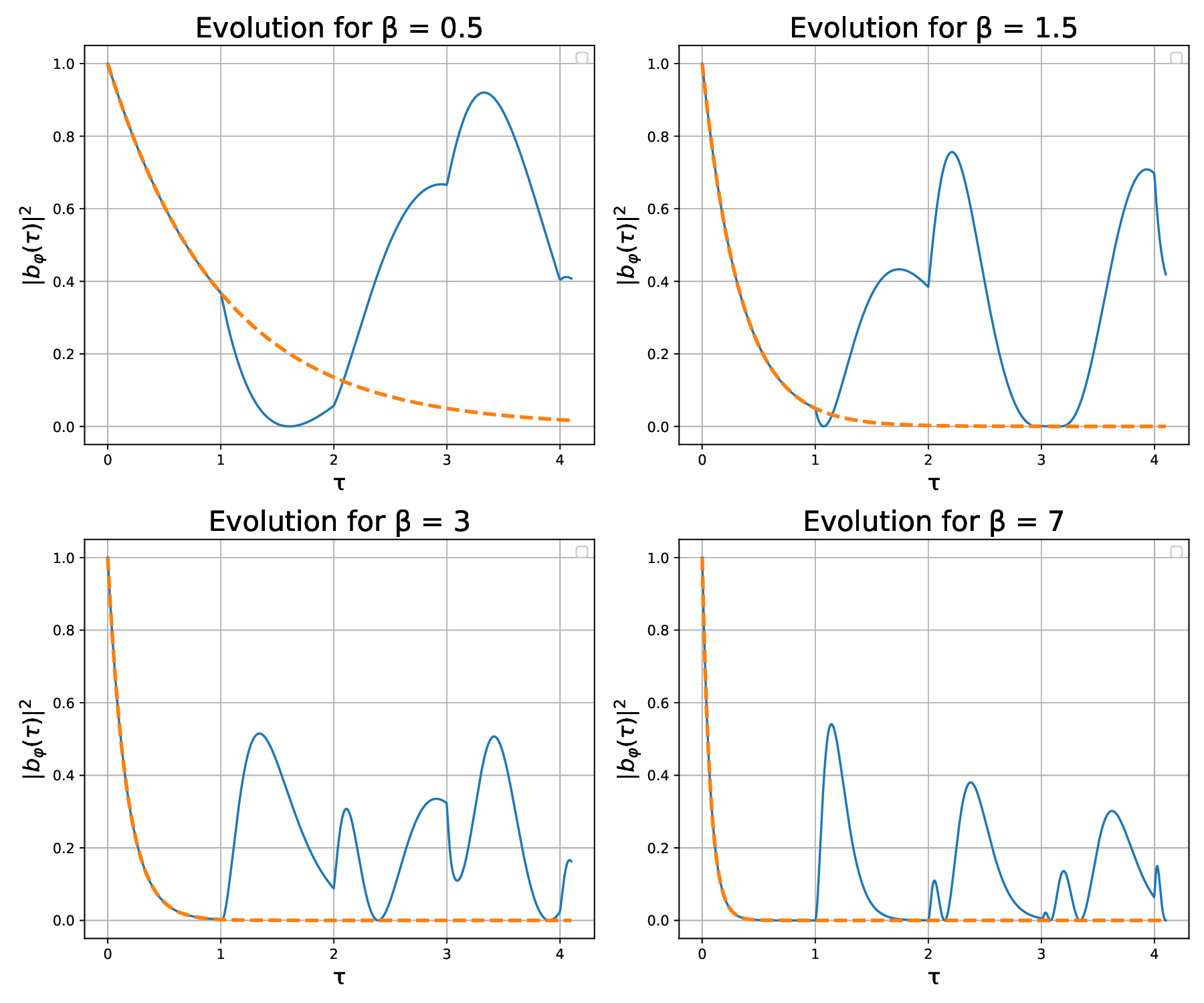}
    \caption{The time evolution of the
    probability $|b_{\varphi}(\tau)|^2$ of the discrete state as a function of dimensionless time
    $\tau$, starting with the initial condition
    $b_{\varphi}(0)=1$, with $\omega_\varphi=0$. The four plots correspond to different values of the 
    dimensionless parameter $\beta$ given by \eqref{eq:beta-defn}. In each case, a dashed orange
    curve represents the exponential decay
    of the Markovian approximation, the decay rate being
    given by the Fermi golden rule.}
    \label{dynamics-plots}
\end{figure}

In Figure \ref{dynamics-plots}, we show a few plots of the decay-revival dynamics
of the Bixon-Jortner system described by the solution \eqref{eq:our-form-solution}.


\subsection{Numerical verification and ladder-state dynamics}\label{sec:numerics}

\begin{figure}[htp]
    \centering
    \includegraphics[width=1.0\textwidth]{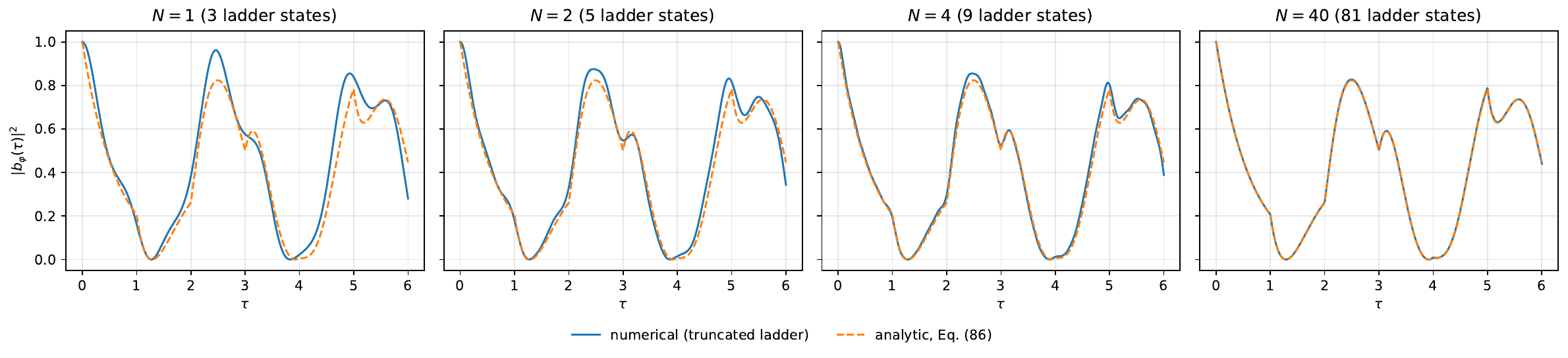}
    \caption{Comparison of the analytic survival probability
    $|b_\varphi(\tau)|^2$ from \eqref{eq:our-form-solution} (orange, dashed)
    with the direct numerical evolution of a truncated Bixon-Jortner
    Hamiltonian with ladder states $n=-N,\ldots,N$ (blue, solid), for
    $N=1, 2, 4, 40$, with $\beta \approx 0.79$ ($v=0.2$, $\Delta=1$) and
    $\omega_\varphi=0$. The numerical result converges to the analytic one
    as the ladder is enlarged, including the non-smooth features at integer
    values of $\tau$. The agreement also confirms the sign of the delayed
    terms in \eqref{eq:our-form-solution}: for $N=40$, the maximum deviation
    is $7\times10^{-3}$, 
    whereas reversing the sign of the sum term 
    in \eqref{eq:our-form-solution}, which corresponds to the expression given in~\cite{barnett2002methods}, results in deviations of order unity.}
    \label{fig:numerical-convergence}
\end{figure}

\begin{figure}[htp]
    \centering
    \includegraphics[width=0.8\textwidth]{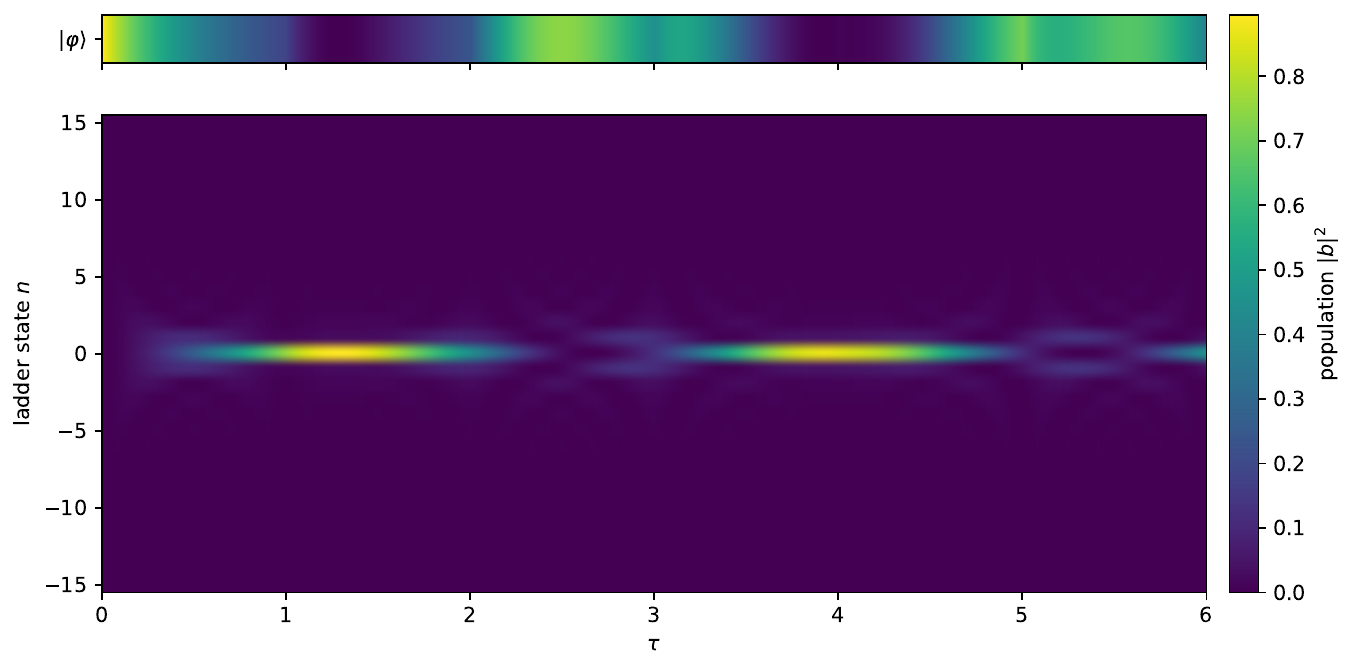}
    \caption{Populations $|b_n(\tau)|^2$ of the ladder states ($|n|\le 15$ shown) and of the discrete state $|b_\varphi(\tau)|^2$ (top
    strip), obtained by direct numerical
    evolution with $N=40$ and the parameters of
    Figure~\ref{fig:numerical-convergence}. The population eventually concentrates in the states close to resonance ($n\approx0$).
    }
    \label{fig:ladder-heatmap}
\end{figure}

\begin{figure}[htp]
    \centering
    \includegraphics[width=0.8\textwidth]{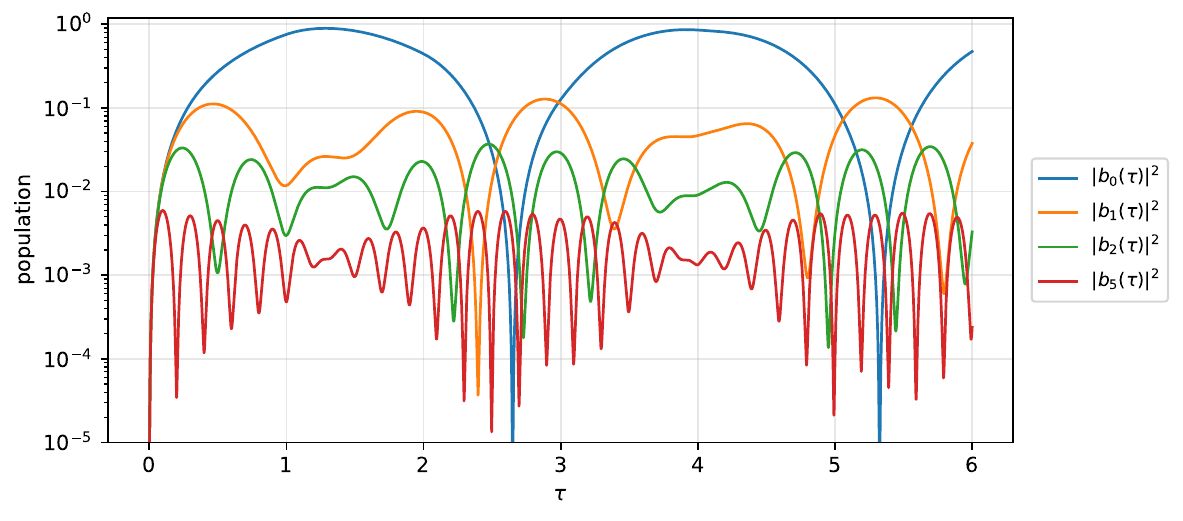}
    \caption{Populations $|b_n(\tau)|^2$ of a few representative ladder states,
    $n=0, 1, 2, 5$, for the run of Figure~\ref{fig:ladder-heatmap}, on a
    logarithmic scale. Since $\omega_\varphi=0$, the populations of the
    states $n$ and $-n$ coincide. The resonant state $n=0$ carries most of
    the population lost by the discrete state, and the populations of the
    more strongly detuned states are smaller on average.}
    \label{fig:ladder-states}
\end{figure}

\begin{figure}[htp]
    \centering
    \includegraphics[width=0.8\textwidth]{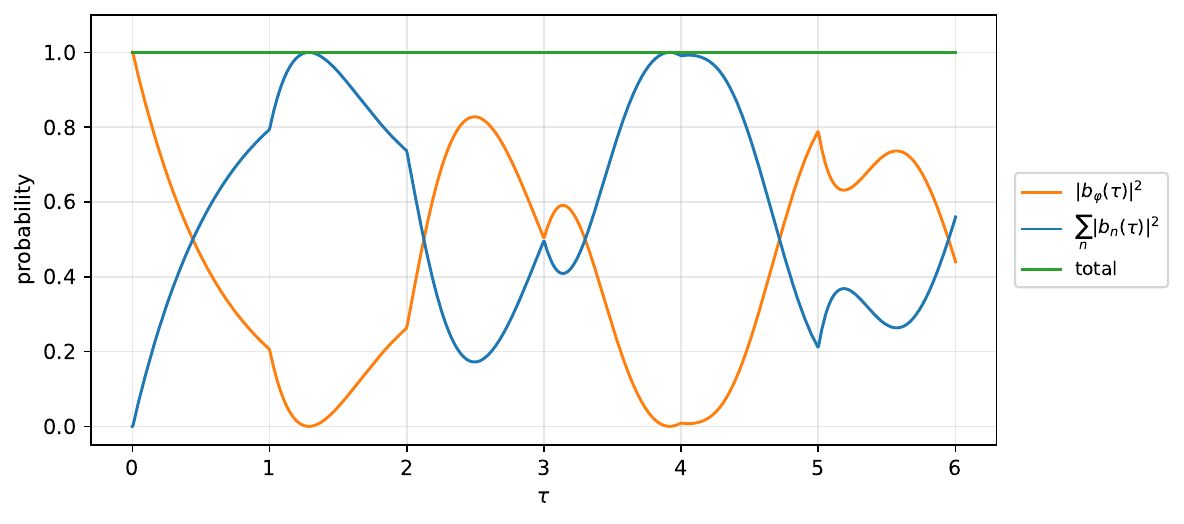}
    \caption{Probability conservation in the numerical evolution of
    Figure~\ref{fig:ladder-heatmap}: the discrete-state population
    $|b_\varphi(\tau)|^2$ (orange), the total ladder population
    $\sum_n |b_n(\tau)|^2$ (blue), and their sum (green), which remains
    equal to 1 to within $2\times10^{-15}$ throughout the evolution.}
    \label{fig:prob-conservation}
\end{figure}

Figures~\ref{fig:example-evolution}
and~\ref{dynamics-plots} were obtained by evaluating the analytic
solution~\eqref{eq:our-form-solution}.
To test the solution independently,
we numerically solved the dynamics of a truncated Bixon-Jortner
Hamiltonian containing the discrete state and the $2N+1$ ladder states $n=-N,\ldots,N$, 
with the discrete state as the initial state. Since the truncated Hamiltonian is a
finite matrix, its time evolution can be computed exactly by numerical
diagonalization without any reference to the integrodifferential or delay
equations derived above. A similar finite-ladder approach, based on the
direct solution of the coupled equations of motion, was used
in~\cite{lavine2023survival} to study the survival probability and the
ladder-state populations of the Bixon-Jortner model.

Figure~\ref{fig:numerical-convergence} compares the numerically obtained
survival probability with the analytic result for increasing ladder sizes.
The truncation is symmetric about the discrete state, and thus corresponds
to the regularization by a symmetric cutoff discussed in
Section~\ref{sec:spectral-density}. For small $N$, the non-smooth features
at integer $\tau$ are smoothed out, in line with the finite width of the cutoff correlation function $C_N(t)$. As $N$ increases, the
numerical survival probability converges to the analytic
one~\eqref{eq:our-form-solution}, which is recovered in the limit
$N\to\infty$. 

The numerical evolution also gives the amplitudes $b_n$ of the ladder
states, which are determined by $b_\varphi$ through
\eqref{eqn:bn-sch}. See Figures~\ref{fig:ladder-heatmap}
and~\ref{fig:ladder-states} for the time evolution of the ladder
state populations.
     At very short times, the discrete state couples with equal strength to all
     ladder states, and the population leaving it is spread over a broad range of
     ladder states. As the phases associated with the detunings
     $n\Delta-\omega_\varphi$ accumulate, the ladder population becomes
     concentrated in a small group of states near resonance---not only the
     resonant state $n=0$, but also its neighbors---which subsequently
     exchange population with the discrete state and with each other in a
     structured, oscillatory manner; for the parameters of
     Figure~\ref{fig:ladder-heatmap}, the states with $|n|\le2$ carry, on
     average, about $90\%$ of the ladder population for $\tau\gtrsim0.5$. This
     structure reflects the fact that the initial state $|\varphi\rangle$
     has most of its weight on a few eigenstates of the full Hamiltonian, i.e., the
     dressed states mentioned in the Introduction. These features are insensitive
     to the truncation: increasing $N$ from $40$ to $160$ changes the ladder
     populations shown by less than $2\times10^{-2}$. A detailed analysis of the
     ladder-state dynamics is beyond the scope of the present work.

Finally, Figure~\ref{fig:prob-conservation}
confirms that the total probability is conserved in the numerical evolution
to machine precision.

\section{Discussion}\label{sec:discussion}

It is well known that the reduced density matrix of one part of a bipartite quantum
system does not in general satisfy a Markovian equation. In certain limits, such a
subsystem satisfies a Lindblad master equation with a time-independent generator,
which \emph{is} Markovian, but in general the time evolution will involve
non-Markovianity due to the part of the total system that is traced over.
Systematic approaches such as the time-convolutionless (TCL) master equation and
the Nakajima-Zwanzig (NZ) equation give corrections to the Markovian behavior.

In this paper, we considered the Bixon-Jortner system, whose Hilbert space is the
direct sum of a discrete state and a ladder, and which, as shown in
Section~\ref{sec:tensor-embedding}, coincides with the one-excitation sector of a
qubit coupled to a bosonic bath with a frequency-comb spectral density. In this
setting, we derived the exact equation of motion \eqref{eq:non-markov-ode} for the
discrete-state amplitude, which takes the very specific (and arguably clean) form of
a delay differential equation. The memory effects are completely explicit, appearing
as updates to the equation at integer times that make the right hand side depend on
new time-delayed copies of the function of interest. Through the tensor-product
embedding, the same amplitude fully determines the reduced dynamics of the qubit as
an amplitude-damping channel, and the revival intervals, where $|b_\varphi|$
increases, are precisely the intervals where this dynamics fails to be CP-divisible.

While delay differential equations are in general difficult to deal with, in this
case we were able to derive the exact solution using a simple ansatz, showing that
the integer-time delays in the equation result in new additive pieces
of the solution at each integer time. In the first time interval, the solution takes
the form of an exponential decay whose rate is given by the Fermi golden rule; in the
open-system language of Section~\ref{sec:tensor-embedding}, the reduced qubit
dynamics here is an amplitude-damping semigroup with constant rate
$\gamma(t)=\Gamma>0$, and hence Markovian.\footnote{Incidentally, this period
corresponds to the limit of the system where the spacing of the ladder goes to zero
with $v^2/\Delta$ kept fixed, making the ladder approach a true continuum, in which
case the exact solution given by Wigner and Weisskopf is recovered
\cite{weisskopf1930berechnung}. See \cite{barnett2002methods} and
\cite{cohen1998atom} for details.}

Since the embedded model is a genuine system-bath model with exactly known reduced
dynamics, it also provides a concrete test case for the general formalisms mentioned
above: the time-local master equation \eqref{eqn:time-local-master}, with rates given
by \eqref{eqn:gamma-s-for-time-local}, is the exact TCL generator of this model,
while the corresponding NZ memory kernel is the temporal Dirac comb
\eqref{eq:corr-poisson}. The delay structure thus makes visible in the time domain
why the TCL rates, initially constant, can turn negative after the first recurrence
time.

The Bixon-Jortner model was originally introduced to describe intramolecular
radiationless transitions~\cite{bixon1968intramolecular}, where the ladder
represents a dense vibronic manifold; there, the recurrence time
$t_0 = 2\pi\hbar\rho$, set by the density of states $\rho$, is typically long.
Through the embedding of Section~\ref{sec:tensor-embedding}, a more direct
realization is a two-level emitter coupled near resonance to the equally
spaced longitudinal modes of a long resonator, whose free spectral range plays
the role of $\Delta$. In this setting, $t_0$ is the round-trip time of the
resonator, the kicks at integer $\tau$ correspond to the return of the emitted
field to the emitter, and the dimensionless parameter $\beta=\Gamma t_0/2$
compares the round-trip time with the golden-rule decay time. Such pulsed,
round-trip-time-structured dynamics was predicted in~\cite{krimer2014route},
and the corresponding multimode regime, with coupling comparable to the free
spectral range, has been reached with a superconducting qubit coupled to a
long transmission-line resonator~\cite{sundaresan2015beyond}.
In a closely related setting, delayed revivals
have been observed directly in the time domain for a superconducting ``giant
atom'' coupled to surface acoustic waves at two distant points, by monitoring
the excited-state population after an initial
excitation~\cite{andersson2019nonexponential}. The finite bandwidth of any
physical resonator provides a natural realization of the regularized coupling
discussed in Section~\ref{sec:spectral-density}.

The model studied here involves a number of idealizations, and it would be
interesting to generalize the present approach beyond them. One natural
direction is a non-constant coupling of the discrete state to the ladder;
in~\cite{isgorur2026coupling}, we study a Lorentzian coupling profile, which
regularizes the model and interpolates between Rabi oscillations and the
decay-revival dynamics of the BJ system, and it would be interesting to
obtain a time-domain solution of this system as well---this work is in progress. 
Other directions include
initial states with more than one excitation or thermal baths, and
interactions beyond the rotating-wave approximation. Counter-rotating terms
break excitation-number conservation, so that the reduction to a
single-excitation sector, and with it the exact delay equation, no longer
holds. In the regime where the rotating-wave approximation is justified
($v,\Delta\ll\omega_0$), however, their effect is expected to be a small
correction that leaves the qualitative features of the single-excitation
dynamics intact, as has been found for the non-Markovian decoherence of a
two-level system in the spin-boson model~\cite{liu2013anomalous}. Finally,
delay differential equations closely related to \eqref{eq:non-markov-ode}
arise for emitters coupled to one-dimensional waveguides at distant points or
in front of a mirror~\cite{guo2017giant,li2024nonmarkovian}, where the delay
is set by the propagation time of the emitted field between coupling points.
Such systems involve a single delay time, whereas in the Bixon-Jortner model
the comb spectrum produces delayed contributions at all integer multiples of
$t_0$, and it would be interesting to explore whether the generating-function
approach used here extends to these settings.

\section*{\textcolor{red} {Acknowledgment and/or disclaimers, if any}}
The authors would like to thank Prof. Stephen M. Barnett for helpful discussions regarding the 
dynamics of the Bixon--Jortner discrete state.
A.Ö. would like to thank the late Metin Sabancıoğlu, \emph{Metin baba}, whose
peaceful home, sense of humor, and unwavering positive attitude toward life provided
the environment where this project first took root.






\section*{\textcolor{red}{Funding, if any}}
This research was supported in part by Boğaziçi University BAP Program under project number 20404 (project code: 25B03D3).

\bibliographystyle{unsrt}
\bibliography{sample}

\end{document}